\documentclass[doublecol]{epl2}

\usepackage{bm}
\usepackage{epsfig}

\newcommand{\av}[1]{\left\langle #1\right\rangle}
\newcommand{\Peff}{\vec{P}_{\rm eff}}
\newcommand{\Ceff}{C_{\rm eff}}

\title{Electron Transport Properties of Composite Ferroelectrics}
\shorttitle{Title} 

\author{O. G. Udalov\inst{1,2} \and A. Glatz\inst{3,4} \and I.~S.~Beloborodov\inst{1}}
\shortauthor{F. Author \etal}

\institute{
  \inst{1} Department of Physics and Astronomy, California State University Northridge - Northridge, CA 91330, USA\\
  \inst{2} Institute for Physics of Microstructures, Russian Academy of Science - Nizhny Novgorod, 603950, Russia \\
  \inst{3} Materials Science Division, Argonne National Laboratory - Argonne, Illinois 60439, USA \\
  \inst{4} Department of Physics, Northern Illinois University - DeKalb, Illinois 60115, USA
}


\abstract{
We study electron transport in composite ferroelectrics --- materials consisting of
metallic grains embedded in a ferroelectric matrix. Due to its complex tunable morphology the
thermodynamic properties of these materials can be essentially different from bulk or
thin-film ferroelectrics. We calculate the conductivity of composite ferroelectrics by taking into
account the interplay between charge localization, multiple grain boundaries, strong Coulomb repulsion,
and ferroelectric order parameter. We show that the ferroelectricity plays a crucial role on the
temperature behavior of the conductivity in the vicinity of the ferroelectric-paraelectric transition.}

\begin{document}

\maketitle

\section{Introduction}

Great efforts in contemporary materials science research focus on properties of composite materials. The interest is motivated by the promise to create materials with unique electrical~\cite{Klin2009,Bon2012}, magnetic~\cite{Brat2011,Nik2002}, thermoelectric~\cite{Bel2009}, optical~\cite{Zhang2011,Zor2010,Ste2008} and elastic~\cite{Wiec2007} properties. The ease of adjusting the electronic structure of composite materials is one of their most attractive assets for fundamental studies of disordered solids and for targeted applications in nanotechnology~\cite{Bel2007review}. Possible applications range from tunable capacitors to ferroelectric tunnel junctions showing giant electroresistance switching effects.
Composite materials are described as solids consisting of normal metallic~\cite{Vinokur2005}, superconducting~\cite{Deut1983,Vill2004,Efetov2002}, or ferromagnetic grains~\cite{Bel2007,Cir2012} embedded into a dielectric matrix.

\begin{figure}
\onefigure[width=0.95\columnwidth]{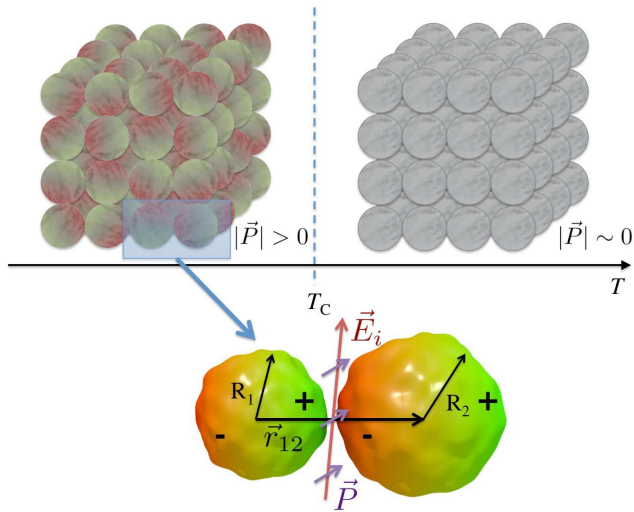}
\caption{(color online). Idealized sketch of a normal metallic granular array embedded in a ferroelectric (FE) matrix below and above the transition temperature in zero external electric field. The zoom shows two of the grains with radii $R_1$ and $R_2$ separated by a layer of FE (not shown) with vector $\vec{r}_{12}$ connecting the grain centers. The vector $\vec{P}$ is the local electric polarization of the FE matrix and the vector $\vec{E}_{i}$ stands for the internal electric field appearing in the system due to the presence of
charge traps in the FE matrix. In general $\vec{E}_{i}$ and $\vec{P}$ are oriented randomly with respect to the vector $\vec{r}_{12}$. The color gradients indicate the induced charge distribution due to the FE material below $T_C$, when the local spontaneous polarization has a finite value. Above $T_C$ the average magnitude of local polarization vanishes.}
\label{Fig_1}
\end{figure}

In this paper we study the electron transport in composite materials consisting of metallic grains embedded into a ferroelectric (FE) matrix in the vicinity of the phase transition. Such materials attract much attention since their possible application in microelectronic devices, for example, in memory cells~\cite{Liu2011,Dai2013}. A high dielectric permittivity makes these materials prominent candidates for capacitor applications~\cite{Tuan2007,Moya2001}. From the point of view of these applications, the study of the electron transport in composite ferroelectrics is a fundamentally important issue.

Transport properties of materials with electric order parameter are well studied and the
conductivity of ferroelectric semiconductors is well known~\cite{FridBook}. In the vicinity of the paraelectric-ferroelectric phase transition, ferroelectric semiconductor conductivity behaves peculiarly due to variations in the ferroelectric semiconductor bandgap and electron scattering by order parameter fluctuations.

In contrast to the situation of ferroelectric semiconductors, we study the case where the conductivity of the ferroelectric
matrix is negligible: there are no electrons in the conduction band and all current carriers
are localized in the metallic grains. In this situation electron cotunneling and
variable range hopping is the only transport mechanism.

Electron tunneling transport properties through single FE barrier are also studied~\cite{Kohl2006,Feng2012,Waser2005,Alexe2010,Tsy2007}, showing the electro-resistance effect. However, coherent multi-grain processes were not studied before.
Therefore, much less is known about electron transport in granular ferroelectrics, when one has to take into account the complex interplay of
Coulomb interaction, ferroelectric ordering, and many grain boundaries in disordered networks of FE barriers.
This defines an urgent quest for a quantitative description of properties of composite FEs.

We note, that granular ferroelectrics were experimentally studied in Refs.~\cite{Cheong2004,Ghosh2006}. In particular, in Ref.~\cite{Ghosh2006}
the results for the metal-insulator transition in granular ferroelectric are shown.
However, most of the published data was obtained in the metallic regime, but the insulating regime is not well characterize yet. Here
we investigate the properties of composite ferroelectrics below the percolation threshold. In this case granular
ferroelectrics are almost insulators.

\section{System description}

Here, we  use the polarization vector, $\vec{P}$, as the ferroelectric order parameter to describe the properties of the FE matrix. For an inhomogeneous medium, $\vec{P}$ has a spatial dependence.
In addition, due to electrostatic disorder, there is an inhomogeneous microscopic electric field $\vec{E}_{i}(\vec{r})$ in the granular medium.

Granularity introduces additional energy parameters into the problem~\cite{Bel2007review}: each nanoscale cluster is characterized by (i) the charging energy $E_c=e^2/(\kappa a)$, where $e$ is the electron charge, $\kappa$ the dielectric constant, and $a$ the granule size, and (ii) the mean energy level spacing $\delta$. The charging energy associated with nanoscale grains can be as large as several hundred Kelvins and we require that $E_c/\delta \gg 1$. This condition defines the lower limit for the grain size: $a_l = (\kappa /e^2\nu)^{1/(D-1)}$, where $\nu$ is the total density of states at the Fermi surface (DOS) and $D$ the grain dimensionality.

The internal conductance of a metallic grain is much larger than the inter-grain tunneling  conductance, which is a standard condition for granularity.
The tunneling conductance is the main parameter that controls the macroscopic transport properties of the sample~\cite{Bel2007review}.
The interplay of local electric field, $\vec{E}_{i}(\vec{r})$, and polarization, $\vec{P}(\vec{r})$, has a profound effect on the tunneling conductance.
This can be understood as follows: If one considers two grains of different sizes (see Fig.~\ref{Fig_1}), the system is characterized by three vectors: i) the microscopic electric field in the region of the grains; ii) the polarization of the ferroelectric matrix near the grains; and iii) the vector connecting the grains $\vec{r}_{12}$. Phenomenologically, the tunneling conductance can be written in the form
$\tilde g_{t}(\vec{P})=g^{0}_{t}(1+\zeta (\vec{E}_{i}\cdot\vec{P})+\mu (\vec{E}_{i}\cdot\vec{r}_{12})(\vec{P}\cdot\vec{r}_{12})+ \gamma(\vec{r}_{12}\cdot\vec{E}_{i})+\epsilon (\vec{r}_{12}\cdot\vec{P})+\tau (\vec{r}_{12}\cdot[\vec{E}_{i}\times\vec{P}]))$,
where $g_t^0$ is the tunneling conductance in the paraelectric state and $\zeta$, $\mu$, $\gamma$, $\epsilon$, $\tau$ are phenomenological constants. Below we consider the case of identical grains. In this situation the system of two grains obeys a spatial inversion symmetry, and correspondingly the coefficients $\epsilon$, $\gamma$ and $\tau$ are zero. Thus the tunneling conductance is given by $\tilde{g}_{t}(\vec{P})=g^{0}_{t}(1 + \zeta (\vec{E}_{i}\cdot\vec{P}) + \mu (\vec{E}_{i}\cdot\vec{r}_{12})(\vec{P}\cdot\vec{r}_{12}))$.
The last term proportional to $\mu$ is finite, since in general $\vec{r}_{12}$ is neither perpendicular to $\vec{P}$ nor to $\vec{E}_{i}$.
The microscopic origin of these contributions is related to the redistribution of the electron density inside the grains due to joint influence of the microscopic electric field and ferroelectric polarization (see Fig.~\ref{Fig_1}).

We introduce the average tunneling conductance
\begin{eqnarray}
g_{t}(P) &\equiv & \av{\tilde{g}_{t}(\vec{P})} =g^{0}_{t}\left(1 + \Ceff \right)\,\,\,\textrm{with}\label{Eq_1} \\
\Ceff &\equiv & \av{\vec{E}_{i}\cdot\Peff}\label{eq.CF}
\end{eqnarray}
being the correlation function of the effective polarization, $\Peff = \zeta \vec{P}+\mu \vec{r}_{12}(\vec{P}\cdot\vec{r}_{12})$ and the electric field $\vec{E}_{i}$.
The average is taken over all pairs of neighboring grains.
Below we first discuss the Ohmic transport in composite ferroelectrics and then summarize our results
for the resistivity in the non-Ohmic regime.

\section{Ohmic transport} There are two distinct mechanisms for electron propagation trough granular arrays at temperatures $T$ below the Coulomb energy, $T < E_c$: elastic and inelastic cotunneling. The essence of a cotunneling process is that an electron tunnels via virtual states in intermediate granules thus bypassing the Coulomb barrier. This can be visualized as coherent superposition of two events: tunneling of the electron into a granule and the simultaneous escape of another electron from the same granule. Elastic cotunneling means that the electron leaving the grain has the same energy as the incoming one. In the event of inelastic cotunneling, the electron coming out of the grain has a different energy than the entering electron. This energy difference is absorbed by an electron-hole excitation inside
the grain, which is left behind in the course of the inelastic cotunneling process. Both mechanisms lead to the following expression
for the conductivity
\begin{eqnarray} \label{Eq_121}
\sigma_{L} &=& g^{0}_{t}\left(1 + \Ceff\right)\exp\left(-\sqrt{T^L_0/T}\right).
\end{eqnarray}
Here $T^L_0$ is the characteristic temperature scale
\begin{eqnarray} \label{Eq_3}
T^L_0 = T_0 \left[1 - \frac{\xi_0}{2a} \ln\left(1 +  \Ceff \right) \right],
\end{eqnarray}
with $T_0=e^2/(k\xi_0)$ and $\xi_0$ being the elastic (inelastic) localization length in the limit of zero polarization~\cite{Bel2007,Bel2007review,Sh1975}
\begin{equation} \label{Eq_4}
\xi^{\rm el}_{0}=a/\ln(E_c/g^{0}_{t}\delta), \hspace{0.5cm}
\xi^{\rm in}_{0}=a/\ln(E^{2}_{c}/T^2g^{0}_{t}).
\end{equation}
It follows from Eq.~(\ref{Eq_121}) that for an uncorrelated microscopic electric field, $\vec{E}_{i}$, and polarization, $\vec{P}$,
the contribution to the Ohmic conductivity due to the ferroelectric order parameter vanishes. However, in any realistic
FE material, the vectors $\vec{E}_{i}$ and $\vec{P}$ are correlated since the local polarization depends on the
grains and on the arrangement of charged impurities.

For small polarization factors, $\Ceff \ll 1$,
the energy scale $T^L_0$ in Eq.~(\ref{Eq_121}), can be written as
$T^L_0 = T_0\left[1-(\xi_0/2a)\Ceff\right]$. Therefore we find for the conductivity of a weakly FE composite material
\begin{eqnarray} \label{Eq_5}
\sigma_{L}^{(0)} = \sigma_{0}\left[ 1+\Ceff\left(1 - \frac{\xi_0}{4a}\sqrt{T_0/T}\right)\right],
\end{eqnarray}
where $\sigma_{0}=g^{0}_{t}\exp(-\sqrt{T_0/T})$ is the conductivity in the paraelectric phase.

\section{Weak non-linear transport}  In the case of a weak external electric field, $E_{e} \ll T/(e\xi_0)$, the corresponding conductivity $\sigma_{W}$ and characteristic temperature $T_0^W$ is obtained using Eqs. (\ref{Eq_121}) and (\ref{Eq_3}), respectively, by replacing $\Ceff\to\Ceff^{(e)}=\av{(\vec{E}_{i}+\vec{E}_{e})\cdot\Peff}$, see Eq.~(\ref{eq.CF}).

\section{Strong non-linear transport} For external electric fields satisfying the condition $T/(e\xi_0) \ll E_{e} \ll \sqrt{E_c\delta}/(ea)$
the transport in granular materials is changing~\cite{Bel2007review}. The last inequality means that the
optimal hopping length is larger than the size of a single grain, $a$, while the first inequality ensures that the
electric field $E_{e}$ is still strong enough to cause non-Ohmic
behavior. In this case the non-linear current is given by the expression
\begin{equation}\label{11}
j=j_0\exp(-(E^W_0/E_{e})^{1/2}),
\end{equation}
where $E^W_0=T^W_0/e\xi_0$ is the characteristic electric field with temperature $T^W_0$ and
$E_e=|\vec{E}_e|$ is the magnitude of the external field \cite{Sh1973}. For small effective polarization we find
\begin{equation}\label{111}
j=j_0 \exp{\left(-\sqrt{\frac{E_0}{E_{e}}}\right)}\left(1-\frac{\Ceff^{(e)}}{4a}\sqrt{\frac{T_0\xi_0}{eE_e}}\right),
\end{equation}
where $E_0=T_0/e\xi_0$ is the characteristic electric field with temperature $T_0$. Equation~(\ref{111}) is valid for fields $E_{e} \gg T/(e\xi_0)$. Therefore, the second term in the brackets is much smaller than the corresponding contribution to the conductivity in the Ohmic regime. This means that the current becomes less dependent on the FE order parameter with increasing external field.

\section{Evaluation of the correlation function} It is clear from our main results, Eqs.~(\ref{Eq_121}) and (\ref{11}), that the conductivity of composite ferroelectrics is determined by the correlation function $\Ceff$ of the internal electric field $\vec{E}_{i}$ and the effective polarization $\Peff$ introduced below Eq.~(\ref{Eq_1}). In general, this correlation function depends on two parameters: i) the external electric field and ii) the temperature.

We first consider the influence of an external electric field. The largest external field in the hopping
regime is $E_{\max}=\sqrt{E_c \delta}/(ea)$. For grain sizes of $5$nm and dielectric constant of order one we find a value of $E_{\max}\sim 10^4$V/cm. The saturation electric field $E_s$ for a typical ferroelectric material is of order of $E_s \sim 10^6-10^7$V/cm~\cite{Frid2006rev}. Therefore,
even for strong non-linear transport the ratio $E_{\max}/E_s$ is very small, $E_{\max}/E_s\sim 10^{-2} \ll 1$, and the influence of
the external electric field on the polarization distribution is negligible, which is used in the following.
Thus, the only parameter controlling $\Ceff$ in Eq.~(\ref{Eq_121}) is the temperature.

We now estimate the internal field $\vec{E}_{i}$ generated by carrier traps, entering the correlation function $\Ceff$ in Eq.~(\ref{Eq_121}). Since metallic grains effectively screen
the electric field, its magnitude between two particular grains is defined by the closest impurity located in the ferroelectric matrix~\cite{Bel2007review}. The magnitude of this field is $E_{i}\sim e/(\kappa r^2) \sim 10^5-10^6$ V/cm with $r$ being
the distance from the closest carrier trap which is of order of a few nm. Thus depending on the concentration of the charged impurities,
the internal field $\vec{E}_{i}$ can be strongly correlated with the polarization $\Peff$. We mention that the impurity ionization energy is several thousand Kelvins and is much larger than the FE transition temperature.
Therefore one can consider the internal field $\vec{E}_{i}$ as temperature independent.

In ferroelectrics at equilibrium, the magnitude and the direction of the order parameter, as well as the ordering temperature $T_C$ depend on many factors such as surface strains~\cite{Pertsev1998,Choi2004,Dawber2005Rew}, depolarization field and its screening~\cite{Bellaiche2004,Bellaiche2007}, ferroelectric thickness or grain size~\cite{Zhao2004,Frid2010rev,Frid2006rev}, and growth conditions~\cite{Bellaiche2006,Wang2009}. The influence of these factors is known for thin ferroelectric films and ferroelectric nanograins. However, less is known for composite ferroelectrics, discussed here.

To describe the polarization behavior of the composite ferroelectrics in the vicinity of ferroelectric transition point
we use Landau-Ginzburg-Devonshire theory~\cite{Tilley2001} with the free energy density written in the form
\begin{equation} \label{Eq_12}
F = F_{0} + \alpha_{1}P^2_{||}+\alpha_{2}\vec{P}^2_{\perp}+ \vec{P}^2(\beta_{1}P^2_{||}+\beta_{2}\vec{P}^2_{\perp})-(\vec{E}_{i}\cdot\vec{P}).
\end{equation}
Here $F_0$ is the free energy independent of the polarization, the vector $\vec{P}$ is the local electric polarization in the region between two particular grains, $P_{||}=\vec{P}\cdot\vec{n}$. The vector $\vec{n}$ describes the uniaxial anisotropy of the system consisting of a thin FE layer confined by grain boundaries.

There are two sources of anisotropy in the system: i) the crystalline anisotropy of the FE, and ii) the FE-metal grains boundaries.
We assume for simplicity that the surface anisotropy is the strongest.
In this case $\vec{n}=\vec{r}_{12}/|\vec{r}_{12}|$ and $\vec{P}_{\perp}=\vec{P}-P_{||}\vec{n}$. The orientation of the
vector $\vec{n}$ is uniformly distributed over the whole solid angle.
We mention that in case of strong enough FE crystal anisotropy the vector $\vec{n}$ has also an uniform angular distribution,
since the FE matrix in granular materials is polycrystalline, and the crystal anisotropy axis varies in space.

A spatial variation of the FE order parameter and transition temperature in between neighboring grains can be neglected. Therefore, Eq.~(\ref{Eq_12}) contains no gradient terms.
The only parameters controlling the polarization behavior are the internal electric field $\vec{E}_{i}$ and the temperature $T$.

Optimizing Eq.~(\ref{Eq_12}) for the free energy we obtain the following equations for the two polarization components
\begin{equation} \label{Eq_13}
\begin{array}{l}
{2\alpha_{1}P_{||}+ 4\beta_{1}P^3_{||}+2(\beta_1+\beta_{2})\vec{P}^2_{\perp}P_{||}-E^{||}_{i}=0},\,\,\textrm{and} \\
{2\alpha_{2}\vec{P}_{\perp}+ 4\beta_2\vec{P}_{\perp}\vec{P}^2_{\perp}+2(\beta_1+\beta_{2})\vec{P}_{\perp}P^2_{||}-\vec{E}^{\perp}_{i}=0.}
\end{array}
\end{equation}
Here $E^{||}_{i}=(\vec{E}_{i}{\cdot}\vec{n})$ and $\vec{E}^{\perp}_{i}=\vec{E}_{i}-E^{||}_{i}\vec{n}$ are the two components of the internal
electric field. Below we consider the isotropic and anisotropic cases separately.

The isotropic model is valid when the internal electric
field gives the largest contribution to the free energy in Eq.~(\ref{Eq_12}) and thus the anisotropy can be neglected.
The anisotropic model is valid in the opposite limit, when the interaction of the local polarization
with the internal electric field is small in comparison with the anisotropy energy.
Any real material is in between these two limiting cases.

For the isotropic system the coefficients in (\ref{Eq_12}) are simplified to $\alpha\equiv\alpha_1=\alpha_2$ and $\beta\equiv\beta_1=\beta_2$. Close to the transition temperature $T_C$ the parameter $\alpha$ has the form $\alpha = \eta(T-T_C)$, and $\beta$ does not depend on temperature. Above the Curie point, $T \gg T_C$ , using Eqs.~(\ref{Eq_13}), we find for the polarization, $\vec{P} = \vec{E}_{i}/(2\alpha)$.
For temperatures $T\ll T_C$ an additional spontaneous contribution emerges, leading to the following result
\begin{equation} \label{Eq_131}
\vec{P}=\sqrt{\frac{|\alpha|}{2\beta}}\frac{\vec{E}_{i}}{|\vec{E}_{i}|}+\frac{\vec{E}_{i}}{4|\alpha|},
\end{equation}
with the first and the second terms being the
spontaneous polarization and the polarization induced by the electric field, respectively.  We mention that the former contribution is much larger than the latter.
We notice that for the isotropic case the electric polarization $\vec{P}$ is directed along the internal electric field $\vec{E}_{i}$
at all temperatures.

Secondly, we consider an anisotropic system with coefficients $\beta_1 \neq 0$ and $\beta_2=0$. In this case, the parameters $\alpha_{1,2}$  can be arbitrary but finite, since for zero $\alpha$ the linear susceptibility would diverge with temperature. For simplicity, we choose $\alpha_1=\alpha_2=\alpha$.  This leads to an isotropic susceptibility for temperatures, $T > T_C$,  and we find for the
polarization $\vec{P} = \vec{E}_{i}/[2\eta(T-T_C)]$. Again, it is parallel to the internal electric field. For low temperatures, $T \ll T_C$,  the anisotropy becomes important, leading to
\begin{equation} \label{Eq_133}
\vec{P}=\sqrt{\frac{|\alpha|}{2\beta_1}}\frac{(\vec{E}_{i}\cdot\vec{n})\vec{n}}{|\vec{E}_{i}|}+\frac{\vec{E}_i}{4|\alpha|}.
\end{equation}
In contrast to the isotropic case, the polarization $\vec{P}$ is directed approximately along the vector $\vec{r}_{12}$
connecting two grains, see Fig.~\ref{Fig_1}.

\begin{figure}
\onefigure[width=0.95\columnwidth]{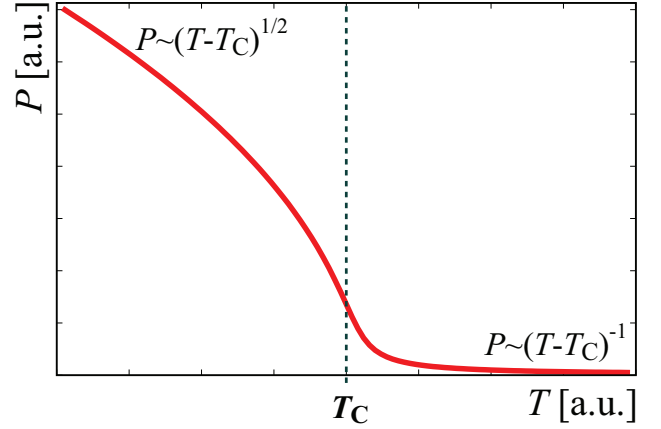}
\caption{(color online). Temperature dependence of the average local polarization $P$ for the isotropic model and zero external
electric field as a function of temperature $T$
(in arbitrary units, $a.u.$).  $T_C$ is the transition temperature to the ordered phase and the asymptotic temperature behavior is given next to the curve below and above $T_C$.}
\label{Fig_2}
\end{figure}

Solving Eqs.~(\ref{Eq_13}) numerically, we find the complete dependence of the local order parameter $\vec{P}$ on temperature for the isotropic model, see Fig.~\ref{Fig_2}. For the anisotropic model this behavior is similar. We mention that for zero external electric field the average polarization of the whole granular system is zero for both models, since the directions of the vectors $\vec{E}_{i}$ and $\vec{n}$ are arbitrary.

Using the above result, we can now calculate the correlation function $\Ceff= \zeta \av{(\vec{E}_{i}{\cdot}\vec{P})} + \mu \av{(\vec{E}_{i}{\cdot}\vec{r}_{12})(\vec{P}{\cdot}\vec{r}_{12})}$ in Eqs.~(\ref{Eq_121}) and (\ref{11}) for the two models for the polarization.
We first estimate the second term in the last expression. Assuming that the vectors $\vec{E}_{i}$, $\vec{P}$, and $\vec{r}_{12}$ follow Gaussian statistics, we write using Wick's theorem $\av{(\vec{E}_{i}{\cdot}\vec{r}_{12})(\vec{P}{\cdot}\vec{r}_{12})} =
 \av{E^{i}_{i}r^{i}_{12}}\av{P^{j}r^{j}_{12}}+\av{E^{i}_{i}P^{j}}\av{r^{i}_{12}r^{j}_{12}}
+ \av{E^{i}_{i}r^{j}_{12}}\av{P^{j}r^{i}_{12}}$.
Here superscripts stand for coordinate indices and summation over repeated indices is assumed.
Since $\vec{E}$ and $\vec{r}_{12}$ are uncorrelated, only the second term is finite and with $\av{r^{i}_{12}r^{j}_{12}} = \delta_{ij} \av{(r^{i}_{12})^{2}}$, we get
$\av{(\vec{E}_{i}{\cdot}\vec{r}_{12})(\vec{P}{\cdot}\vec{r}_{12})}= \av{\vec{r}^{~2}_{12}} \av{(\vec{E}_{i}{\cdot}\vec{P})}$.

We now estimate the correlation function using our results for the polarization. For the isotropic ferroelectric matrix we find
\begin{equation}\label{Eq_144}
\Ceff = \tilde{\zeta}\left\{\begin{array}{l} \sqrt{|\alpha|/(2\beta)}\av{|\vec{E}_{i}|}+\av{\vec{E}^2_{i}}/(4|\alpha|), \hspace{0.2cm} T \ll T_C
\\ { \av{\vec{E}^2_{i}}/(2\alpha), \hspace{3.2cm} T \gg T_C,}
\end{array}\right.
\end{equation}
where $\tilde{\zeta}=\zeta+\mu \av{\vec{r}^{~2}_{12}}$. For the anisotropic model and low temperatures $T < T_C$ we replace $\av{|\vec{E}_{i}|}$ in Eq.~(\ref{Eq_144}) by $\av{|\vec{E}_{i}| \cos^2\phi}$,
where $\phi$ is the angle between vectors $\vec{n}$ and $\vec{E}_{i}$. Since the position of the charge traps and mutual orientation of grains are uncorrelated, the distribution function for the angle $\phi$ is uniform, thus $\Ceff$ is given by Eq.~(\ref{Eq_144}) with the replacement $\beta\to 4\beta_1$.

At this point, we point out that the Ohmic conductivity of composite ferroelectrics in Eq.~(\ref{Eq_121}) in the vicinity of the Curie point depends on temperature through two parameters: i) the correlation function and ii) the localization length $\xi_0$. For temperatures $T < T_{\rm el}=\sqrt{\delta E_c}$, elastic cotunneling is the main mechanism for electron transport with
the localization length $\xi_0$ being independent of temperature~\cite{Bel2007review}. For grain sizes
$\sim 5$nm the temperature $T_{\rm el}$ is of order $40$K. For most FEs the phase transition occurs above room temperature, i.e., $T_C \gg T_{\rm el}$.
Therefore, the electron transport near $T_C$ is due to inelastic cotunneling and we find for the conductivity of the isotropic model in the Ohmic regime the following result
\begin{equation}\label{EQ_15}
\frac{\sigma_L}{\sigma_0} = \left\{\begin{array}{l} {\!1+\tilde{\zeta}\sqrt{\frac{\eta(T_C-T)\av{|\vec{E}_{i}|}^2}{2\beta}}
\left[1 - \frac{\xi_0}{4a}\sqrt{\frac{T_0}{T}}\right], T\ll T_C,}
\\ \!1+\tilde{\zeta}\frac{\av{\vec{E}^2_{i}}}{2\eta(T-T_C)}
\left[1 - \frac{\xi_0}{4a}\sqrt{\frac{T_0}{T}} \right], \hspace{1.3cm} T\gg T_C.
\end{array}\right.
\end{equation}
Here the temperature scale $T_0$ was defined below Eq.~(\ref{Eq_3}).
Equation~(\ref{EQ_15}) is valid for the anisotropic model as well with the substitution $\beta \rightarrow 4\beta_1$.

To find the conductivity in the weak non-linear regime we need to calculate $\Ceff^{(e)}$. For the isotropic model it has the form
\begin{equation}\label{Eq_16}
\Ceff^{(e)} = \tilde{\zeta}\left\{\begin{array}{l} { \sqrt{|\alpha|/(2\beta)}\av{\!|\vec{E}_{i}|\!+\!\vec{E}^2_{e}/(4|\vec{E}_{i}|)\!} , \, T \ll T_C,}
\\  \left[\av{\vec{E}^2_{i}}+\vec{E}^2_{e}\right]/(2\alpha), \hspace{2cm} T \gg T_C.
\end{array}\right.
\end{equation}
It follows from above equations that an external field leads to quadratic corrections of order of $E^2_{e}/E^2_{i} \ll 1$.
To obtain the expression for conductivity in the weak non-linear regime, one should replace $|\vec{E}_{i}|$ in the first line of Eq.~(\ref{EQ_15})
by   $|\vec{E}_{i}| \rightarrow |\vec{E}_{i}|+\vec{E}^2_{e}/(4|\vec{E}_{i}|)$, and  $\vec{E}^2_{i}$ in the second line of Eq.~(\ref{EQ_15}) by $ \vec{E}^2_{i} \rightarrow \vec{E}^2_{i}+\vec{E}^2_{e}$.

In the strong non-linear regime, the current through isotropic composite ferroelectric, Eq.~(\ref{111}), can be written as
\begin{equation}\label{EQ_18}
\frac{j}{J_0}=\left\{\begin{array}{l} {\!1- \frac{\tilde{\zeta}}{4a}\sqrt{\frac{T_0\xi_0}{e E_e}}\sqrt{\frac{\eta(T_C-T)\av{|\vec{E}_{i}|}^2}{2\beta}}, \hspace{0.2cm} T\ll T_C,}
\\ \!1-\frac{\tilde{\zeta}}{8a}\sqrt{\frac{T_0\xi_0}{e E_e}}\frac{\av{\vec{E}^2_{i}}}{\eta(T-T_C)},  \hspace{1.4cm} T\gg T_C.
\end{array}\right.
\end{equation}
Here $J_0=j_0\exp{(-(E_0/E_{e})^{1/2})}$. The conductivity for the anisotropic model can be obtained using the isotropic result by
replacing $\beta$ by $4\beta_1$ in Eqs.~(\ref{Eq_16}) and (\ref{EQ_18}).

\begin{figure}
\onefigure[width=0.95\columnwidth]{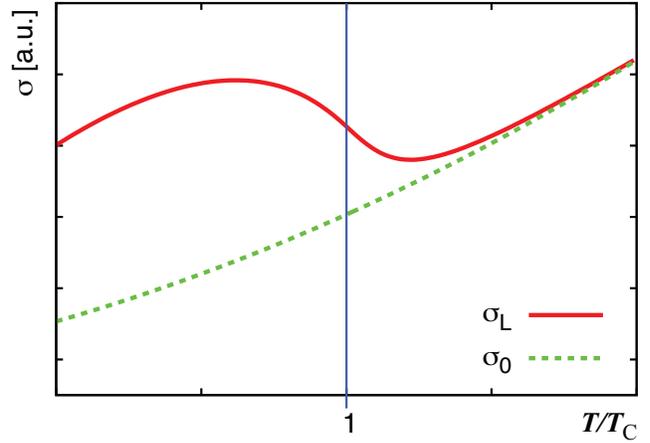}
\caption{(color online). Temperature dependence of the conductivity of granular materials in the Ohmic regime $\sigma_L$, Eq.~(\ref{EQ_15}).
The solid (red) and dashed (green) lines correspond to a granular ferroelectric and a granular metal, respectively. $\sigma_0$ is the conductivity in the paraelectric phase. $T_C$ is the ferroelectric Curie temperature.}
\label{Fig_T}
\end{figure}

Figure~(\ref{Fig_T}) shows the temperature dependence of the conductivity of granular ferroelectrics (solid line) in the
linear response regime, Eq.~(\ref{EQ_15}). This behavior can be understood as follows: i) the increase of conductivity
with temperature appears due to increase in the number of phonons leading to a larger hopping probability.
This feature is related to the intergrain hopping mechanism and does not depend on the FE matrix;

ii) The second factor is the influence of ferroelectric matrix on the hopping probability.
Above the Curie temperature $T > T_C$ the influence of FE matrix on transport is negligible due to small electric polarization.
The polarization grows with decreasing the temperature, see Fig.~(\ref{Fig_2}), leading to larger intergrain conductance.
The interplay between mechanisms i) and ii) leads to the appearance of non-monotonic temperature behavior of conductivity with
some peculiarity in the vicinity of Curie temperature, $T_C$.

The dashed line in Fig.~(\ref{Fig_T}) corresponds to the case of an insulating matrix instead of a FE matrix.
In this case only the first mechanism is important leading to a monotonic increase of the conductivity.

Finally, we discuss the assumptions and applicability of our approach. It is well known that the ferroelectric order
parameter shows hysteretic behavior below the transition temperature, $T_C$, meaning that the FE state depends
on its history. The results derived in this paper assume that the FE matrix is in its ground state.
This is justified for two cases: 1) when both the hysteresis loop and the external field are small enough
in comparison with the internal field or 2) when changes to external parameters are done adiabatically. For strong electric fields,
hysteresis effects need to be taken into account, since these will have an influence on the transport properties.

The applicability of the Landau-Ginzburg-Devonshire theory implied another restriction of our model.
Near the FE Curie temperature, fluctuations of the order parameter, become comparable to the average polarization.
Therefore, our approach is not valid in this region. Estimates for BaTiO$_3$ ($T_C \approx 400 K$) show that this region is less than $1 K$
around the Curie temperature.

\section{Conclusions}
In conclusion, we studied the electron transport in composite ferroelectrics.
Due to the complex morphology and tunability of these materials, targeted applications are possible, exceeding the performance of
bulk and thin-film ferroelectrics.
We calculated the conductivity of composite ferroelectrics, taking into account effects of interference
between charge localization, multiple grain boundaries, strong Coulomb repulsion, and ferroelectric order parameter.
We showed that the FE matrix plays a crucial role on the temperature behavior of the
conductivity in the Ohmic and non-Ohmic regimes.

\acknowledgments
We thank Nikolai Chtchelkatchev and Nick Kioussis for useful discussions.
A.~G. was supported by the U.S. Department of Energy Office of Science under the Contract No. DE-AC02-06CH11357.
I.~B. was supported by NSF under Cooperative Agreement Award EEC-1160504 and NSF PREM Award DMR-1234567.

\bibliographystyle{eplbib}
\bibliography{granule}

\end{document}